\documentclass{aa}  

\usepackage{amsmath}
\usepackage{soul}
\usepackage{tablefootnote}
\usepackage{bm}
\usepackage{xcolor}
\usepackage[normalem]{ulem}
\usepackage[
    range-units=single,         
    range-phrase=-,             
    separate-uncertainty=true,  
    multi-part-units=single     
] {siunitx}
\DeclareSIUnit\angstrom{Å}
\DeclareSIUnit\gauss{G}
\DeclareSIUnit\Ic{I_c}
\definecolor{al}{rgb}{0.6,0.2,0.0}
\definecolor{as}{rgb}{0.8,0.2,0.3}
\definecolor{mvn}{rgb}{0.5,0.5,0.9}
\definecolor{fis}{rgb}{0.82,0.41,0.12}

\newcommand{\fluxunit}{{\rm erg~s^{-1}~cm^{-2}}}

\usepackage{graphicx}
\usepackage{txfonts}

\usepackage[colorlinks,linkcolor=blue,citecolor=blue,linktocpage=true,breaklinks, 
plainpages=false,urlcolor=blue]{hyperref}

\definecolor{iimm}{cmyk}{1.0,0.0,1.0,0.3}
\definecolor{ttkk}{cmyk}{0.57,0.3,0.0,0.0}

\definecolor{deepmagenta}{rgb}{0.8, 0.0, 0.8}

\usepackage{booktabs} 
\usepackage{multirow} 
\usepackage{adjustbox}
\usepackage{caption} 
\usepackage{orcidlink} 
\begin{document}

   \title{Wave excitation by a collapsing granule: insights from IFU observations and high-resolution RMHD simulations}

   \author{J.~Buttler \inst {1} \orcidlink{0009-0002-2135-5575}, G. Vigeesh\inst{1}\orcidlink{0000-0002-9820-9114}, I. Mili\'{c}\inst{1,2,3} \orcidlink{0000-0002-0189-5550}, 
          M. van Noort \inst{4}, S. M. {D\'iaz~Castillo} \inst{1,5} \orcidlink{0000-0002-5330-3131}, C. J. D\'iaz Baso \inst{6,7}\orcidlink{0000-0001-9239-9482}, 
          F. Riva\inst{8}\orcidlink{0000-0002-0264-442X}, 
          O.~Steiner \inst{1,8}\orcidlink{0000-0003-4708-1074}
          }
   \institute{
    Institut f\"{u}r Sonnenphysik (KIS), Georges-K\"ohler-Allee 401A, 79110 Freiburg, Germany
    \and 
    Faculty of Mathematics, University of Belgrade, Studentski Trg 16, 11000 Belgrade, Serbia 
    \and 
    Astronomical Observatory, Volgina 7, 11060 Belgrade, Serbia 
    \and 
    Max-Planck Institute f\"{u}r Sonnensystemforschung, Justus-von-Liebig-Weg 3, 37079 G\"{o}ttingen, Germany 
    \and
    French-Spanish Laboratory for Astrophysics in Canaries (FSLAC), THEMIS S.L.U., Av de los Menceyes 93, 38205 La Laguna, Tenerife, Spain
    \and
    Institute of Theoretical Astrophysics, University of Oslo, P.O. Box 1029 Blindern, N-0315 Oslo, Norway
    \and
    Rosseland Centre for Solar Physics, University of Oslo, P.O. Box 1029 Blindern, N-0315 Oslo, Norway
    \and
    Instituto richerce solari Aldo e Cele Dacc\'{o} (IRSOL), Faculty of Informatics, Universit\`a della Svizzera italiana (USI), 6605 Locarno, Switzerland\\
   \email{jan.buttler@leibniz-kis.de}}
  
   \date{Received ; accepted }

  \abstract
   {Granular collapse is an ubiquitous process of granular evolution on the solar surface, but it is hard to analyze in sufficient spatial, temporal, and spectral detail.}
   {We analyze the change in physical conditions in the photosphere during a specific granular collapse event and the subsequent atmospheric response.}
   {We contrast a high-resolution radiative magneto-hydrodynamic simulation of a granular collapse performed using the CO5BOLD code with the recent integral field unit observations carried out using the MiHI instrument at the Swedish 1-m Solar Telescope.}
   {Our analysis shows that the observed and simulated granular collapse show remarkable similarity. Specifically, they both exhibit the signature of a wave pulse excited in the deep photosphere and visible up to the temperature minimum. This wave is detectable through a blue-wing emission in the observed and synthetic \ion{Na}{i}~D1 line. We also estimate the acoustic energy flux carried by the wave and analyze its initiation.}
   {Combining high-resolution IFU spectropolarimetry and state-of-the-art simulations of the solar lower atmosphere, this study showcases our current capabilities in identifying specific physical processes taking place during the granular collapse and their impact on the atmosphere above.
   }
     \titlerunning{Collapsing Granule in high resolution}
    \authorrunning{Buttler et al.}

   \maketitle
   \nolinenumbers

\section{Introduction}
\label{sec:intro}

The quiet solar surface consists of granules: hot, upflowing plasma elements separated by a network of colder, downflowing, intergranular lanes. They are a direct signature of the convection taking place in the outer layers of the Sun \citep[][]{Nordlund_2009_review}. Granular convection is presumed to be the dominant excitation mechanism for the acoustic-gravity waves that eventually propagate into the chromosphere, possibly contributing to the chromospheric heating \citep{Skartilien_2000_pmodeconv, Carlsson_2019_rev}. The lifetime of the granules is approximately 10 minutes, and they end their lives through the processes of collapse, fragmentation, or merging \citep[e.g.][]{Hirz_1999_II, Mueller_2001_granules, SDC_2022_mlgran}. Granular collapse is characterized by a rapid \citep[a few minutes, ][]{Skartlien2000collapse} change from an upward to a downward granular flow, and an accompanying decrease in temperature. While granular collapse is a purely radiative-hydrodynamic phenomenon \citep{Skartlien2000collapse}, in magnetized regions it can induce either the merging of nearby magnetic bright elements or the convective collapse of magnetic flux tubes \citep[e.g.][]{Parker_1978_collapse, SpuitZ_1979_tubeinstability, Danilovic2010mhd_vs_hinode}. These processes may, in turn, trigger acoustic activity in the higher atmosphere via photospheric rebound flows \citep{Grossmann_D_1998_convcollapse, BellotRubio2001collapse, DiazCastillo2024connectivity}.

While granular collapse is a ubiquitous process in the solar photosphere, there are only a handful of studies that focus on a detailed analysis of specific events. 
Among them, \citet{Nagata_2008_fluxtubeformation} observed downward flows that resulted in the creation of a magnetic bright point using the spectropolarimeter \citep[SP;][]{Lites_2013_SP} attached to the Solar Optical Telescope \citep[SOT;][]{Tsuneta_2008SoPh_SOT} onboard the Hinode mission \citep{Kosugi2007hinode}.
More recently, \citet{2020A&A...642A.154K} observed several cases of wave excitation caused by different granular events, including a collapsing granule, using the Goode Solar Telescope \citep[GST;][]{Cao2010GST}. The wave analysis focused on the chromosphere, using the FISS instrument \citep{Chae_2013_FISS}. Due to the lack of spectropolarimetry, the magnetic fields could not be inferred. These works relied on scanning slit spectrographs, which limited both the cadence and the spatial resolution. 

Because collapsing granules evolve over only a few minutes and occupy a spatial extent of approximately one Mm and less, their study requires an instrument capable of both high spatial and temporal resolution, preferably with a high-fidelity spectral sampling, to constrain the variation of physical parameters with height. Integral field units (IFUs) are particularly well suited for this task, as they provide simultaneous spatial and spectral sampling over a limited field of view. Here we present the first analysis of a collapsing granule observed with the Microlensed Hyperspectral Imager \citep[MiHI,][]{vanNoort2022A&AMihiInstrument} prototype at the Swedish 1-m Solar Telescope \citep[SST,][]{Scharmer2003sst}. The spectral cameras of the MiHI prototype operate at a frame rate of 30\,Hz, which enables image restoration simultaneously with spectral deconvolution \citep[][]{vanNoort2022MiHI3} and post facto cadence selection down to the polarimetric modulation rate of 7.5\,Hz. The observations contain the sodium\,D1 spectral line, which samples the photosphere and temperature minimum, and two weaker photospheric lines of iron and nickel, which provide additional photospheric information. The cadence of this dataset is 2.7\,s, which is an order of magnitude improvement over earlier works. 

We present the detection and analysis of a collapsing granule found in this unique dataset. We use simple and robust velocity and magnetic field diagnostics and flow tracking techniques to characterize the spatiotemporal evolution of physical parameters throughout the collapse. We then compare both the inferred quantities and the observed spectra with a simulated granular collapse obtained with the CO5BOLD code \citep{2012JCoPh.231..919F}. We detected and characterized an upward wave propagation event, initiated by the collapse, and visible as a blue-shifted emission in the \ion{Na}{i} line. This emission is evident both in the simulated and the observed spectra. Analysis of the simulation allowed us to estimate the energy balance of such an event and to characterize its contribution to the upward energy flux. These results provide new insight into the interplay between convection and the magnetic field during a granular collapse event and the role of such events in energy transport throughout the lower solar atmosphere.

In Section~\ref{sec:obs}, we present the MiHI observations and the diagnostic techniques to infer the physical parameters in the observed patch of the atmosphere. Section~\ref{sec:simulations} presents the simulation setup and the calculation of synthetic Stokes spectra from the simulation. In Section~\ref{sec:comparison}, we directly compare the granular collapse found in the simulations and the observations, with an emphasis on a wave event and its spectral signatures, detected in both datasets. We provide additional discussion and conclusions in Section~\ref{sec:conclusions}.

\section{Observations and diagnostics}
\label{sec:obs}

\subsection{Observations}

\begin{figure}[htbp]
    \centering
    \includegraphics[width=0.5\textwidth]{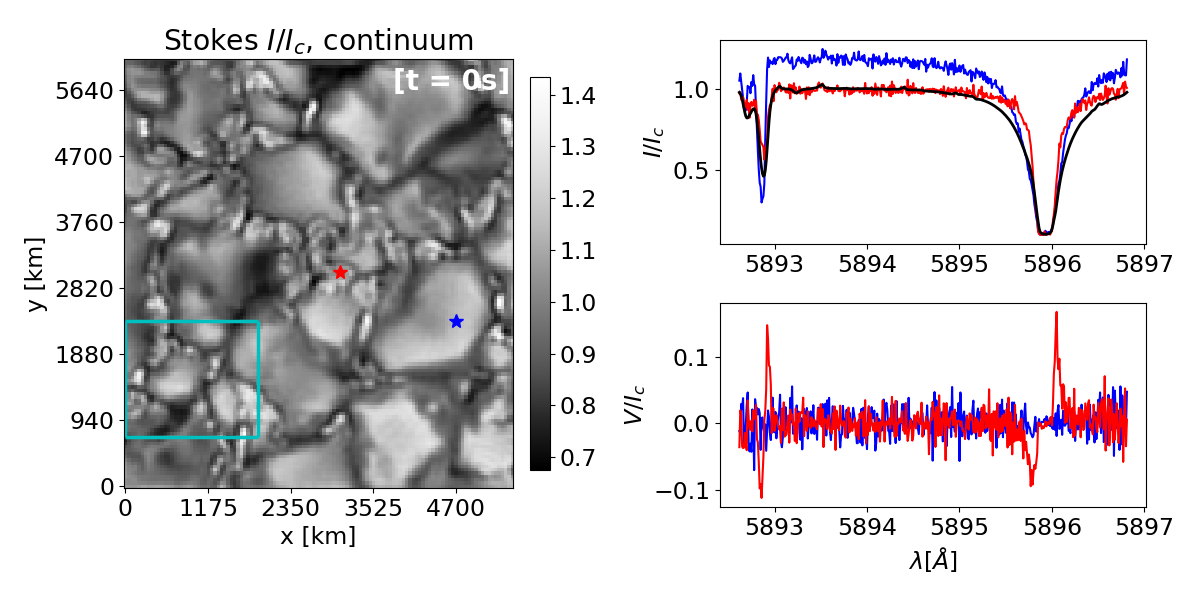}
    \caption{Left panel: Continuum image at $\lambda=589.4$~nm, integrated over 0.01~nm, showing the full MiHI FOV at the beginning of the collapse (09:24~UT). The cyan box marks the granule undergoing the collapse,  which is shown in detail in Fig.\,\ref{fig:inv_evol}. Right panel: Stokes $I$ (top) and $V$ (bottom) for two selected pixels marked with red and blue asterisks in the left panel. The Stokes $I$ spectrum averaged over the whole FOV is shown in black. All the data are normalized to the mean continuum intensity over the MiHI FOV at the given time. Stokes $V$ is integrated over approximately 8 seconds to reduce noise.}  
    \label{fig:fov_outline}
\end{figure}

The observations were taken at the SST on August 17th, 2018, from 08:57:18 to 09:40:05~UT, with the final cadence of approximately 2.7\,s. The field of view (FOV) was centered near disk center at the location $-88'' {\rm E}, 97'' {\rm N}$, resulting in the cosine of the heliocentric angle of $\mu = 0.99$. The FOV of MiHI covers 128 $\times$ 115 pixels with a spatial sampling of 0.065\,arcsec per pixel (47\,km on solar surface), which is close to the critical sampling of 0.062\,arcsec per pixel required for the SST focal plane at these wavelengths. The observing conditions were favorable, yielding restored data close to the diffraction limit. The total observed spectral region extended from 5892.2 to 5897.3~\AA. The following analysis focuses on the $5892.6-5896.8\,{\rm\AA}$ range, which contains the \ion{Na}{i}\,D1 line and two weaker, photospheric lines of \ion{Fe}{i} and \ion{Ni}{i}. For more details on the observed spectral lines, see Table~\ref{tab:lines}. The spectral sampling was 1\,pm (corresponding to a spectral resolution higher than $2\times10^5$), and a spectral deconvolution was performed during the data restoration. The right panel of Fig.~\ref{fig:fov_outline} shows Stokes $I$ and $V$ spectra of two example pixels. 

\begin{table}[]
    \caption{Spectral lines observed by MiHI.}
    \centering
    \begin{tabular}{c|c|c|c}
    Line & $\lambda_0\,$[\AA] & $g_{\rm eff}$ & Formation \\
    \hline
        \ion{Fe}{i} & 5892.69 & 1.83 & LTE, photosphere \\
        \ion{Ni}{i} & 5892.87 & 1.0 & LTE, photosphere \\
        \ion{Na}{i} D1 & 5895.92 & 1.33 & NLTE, temp. minimum \\
    \hline
    \end{tabular}
    \label{tab:lines}
\end{table}

The observations (see the left panel of Fig.~\ref{fig:fov_outline}) cover several granules permeated by a fine pattern of magnetic features \citep[filigree, see e.g.][]{Dun_1973_filigree}. Throughout the observation time span ($\approx 43$~minutes) at least five granular collapse events can be spotted, along with other dynamic phenomena. We focus on the specific collapse event starting at 09:24~UT (from now on referred to as $t=0$). This particular collapse is selected because it was observed under excellent seeing conditions and the granule has a regular, circular shape. A similar event at the beginning of the observations was discarded due to unstable seeing.

The cyan box in the left panel of Fig.~\ref{fig:fov_outline} shows the region where the collapse happens. 

Figure~\ref{fig:inv_evol} shows the gradual collapse of the granule under consideration, which results in the spatial reordering of the magnetic field and the vertical velocity field in the photosphere. The collapse lasts for approximately 8 minutes. This is significantly longer than the simulated and observed collapses reported by \citet{Skartlien2000collapse} and \citet{2002A&A...390..681H}. This difference may be due to the presence of the relatively strong magnetic field, which is shown in the two rightmost panels of Fig.\,\ref{fig:inv_evol}. 

\subsection{Magnetic field and line-of-sight velocity inference}
\label{ssection:magneticfieldinference}

\begin{figure}
    \centering
    \includegraphics[width=1\linewidth]{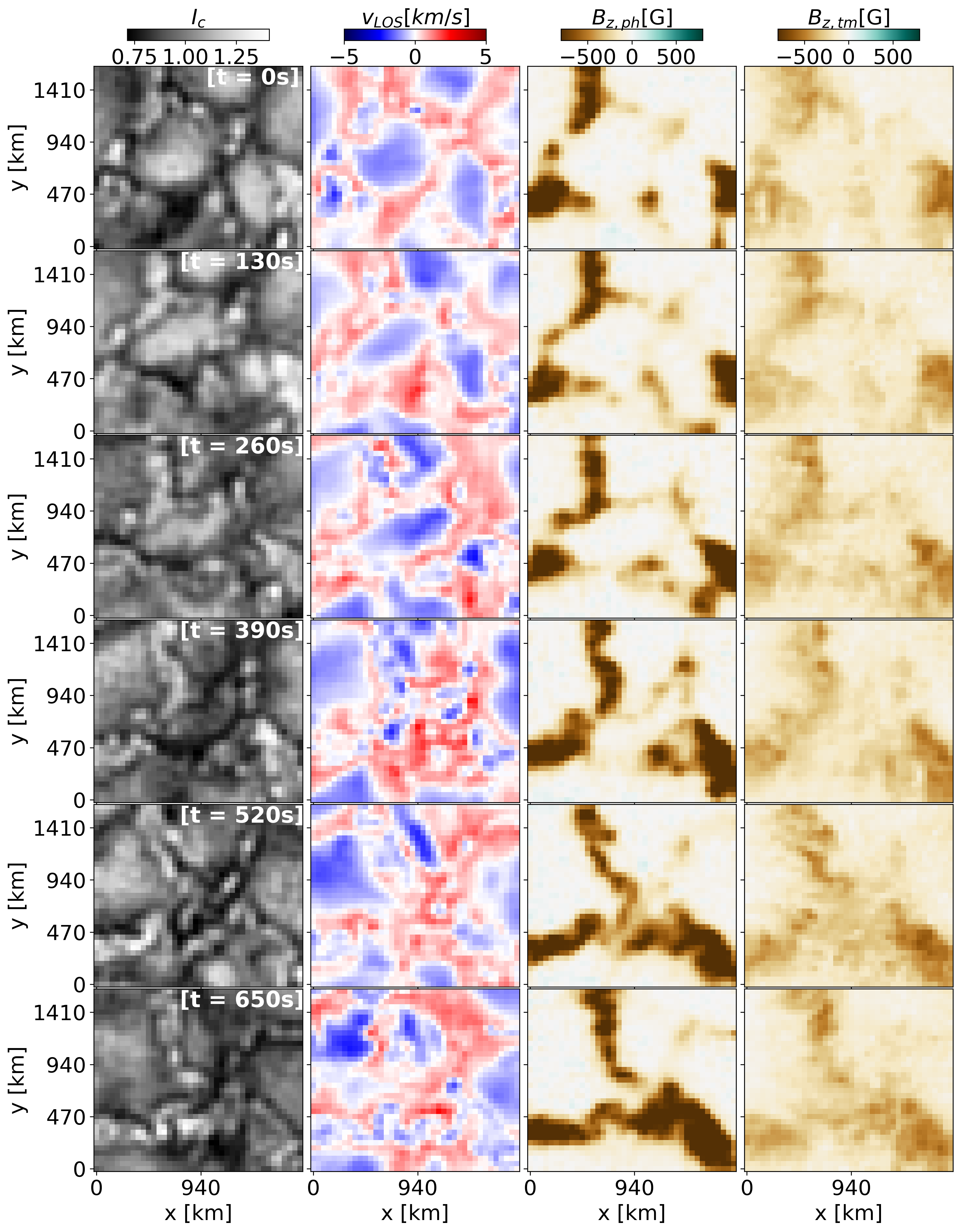}
    \caption{Evolution (from top to bottom) of atmospheric physical parameters during the granular collapse in the observations. From left to right: continuum intensity ($I_c$), photospheric LOS velocity ($v_{LOS}$), photospheric LOS magnetic field ($B_{z,ph}$), temperature minimum LOS magnetic field ($B_{z,tm}$). The average LOS velocity over the whole MiHI FOV was subtracted to remove p-modes and other systematics.}
    \label{fig:inv_evol}
\end{figure}

To interpret the Stokes spectra of the two photospheric lines (\ion{Fe}{i} and \ion{Ni}{i}) and obtain an estimate of the photospheric magnetic field and line-of-sight (LOS) velocity, we used the PyMilne spectropolarimetric inversion code \citep{Jaime2019pm}. This code is based on the Milne-Eddington approximation and supports spatial and temporal regularization \citep[][]{Jaime2024spacetreg}. The use of regularization in the inversion process is critical to interpret MiHI data, which, due to very high spatial resolution, fast-cadence, and spectral resolution, shows a low nominal signal-to-noise ratio \citep[also discussed in][]{vanNoort2022MiHI3}. For the two photospheric lines, we used Tikhonov regularization in the spatial domain and inverted the two lines simultaneously. This is achieved by using the same source function for both lines and calibrating the ratio of their strengths so that the fit to the Stokes $I$ is as good as possible. We chose the spatial regularization parameter $\alpha$ \citep{Jaime2019pm}, by re-doing the inversion with successively increasing value of $\alpha$, and stopping once the $\chi^2$ metric of the fit is $10\%$ higher than in the case without the regularization. This procedure yields a reasonable balance between the quality of the fit to the observed Stokes profiles and the smoothness of the resulting parameter maps in the photospheric quantities, as can be seen from the second and third columns in Fig.~\ref{fig:inv_evol}, so we deemed temporal regularization unnecessary.

To interpret the circular polarization in the \ion{Na}{i} line, we used the temporally regularized weak-field approximation (WFA) of \citet{DiazBaso2025_neuralwfa}\footnote{\url{https://github.com/cdiazbas/neural_wfa}}. We applied the WFA to the wavelength range of $\pm$ 0.1\,\AA\ around the Na~I~D1 line core, thus effectively sampling the layers around the temperature minimum and avoiding the contribution from the lower layers. We opted to use only the temporal Tikhonov regularization, because the introduction of additional spatial regularization resulted in maps that were overly smooth in the temperature minimum region. This combination of approaches, where we use spatially regularized ME inversion in the photosphere and the temporally regularized WFA in the temperature minimum, results in clear $B_z$ maps, both in the photospheric and temperature minimum layers (see Fig.\,\ref{fig:inv_evol}), while the linear polarization signals are still too noisy to allow the inference of the horizontal magnetic field component.

\subsection{Horizontal velocity inference}

We inferred the horizontal velocity field from the continuum intensity evolution using DeepVel \citep{AsensioRamos2017}, a deep learning model based on a fully convolutional neural network architecture that retrieves horizontal velocities from consecutive continuum intensity images. Contrary to the conventional tracking methods, it does not require space- and time-averaging, and it has demonstrated good performance in inferring granular and subgranular components of the plasma flow, compared to other intensity-based reconstruction algorithms \citep{Tremblay2018}. The DeepVel approach has been verified through validation tests on unseen simulated datasets produced by a different code from the one used for training the network \citep{AsensioRamos2017, Lennard_2025_DeepVelARs}.

To apply the DeepVel model to the MiHI data, we trained the model using synthetic intensity maps from a comprehensive MURaM \citep{Vogler2005muram} quiet-Sun radiative magneto-hydrodynamic (rMHD) simulation \citep{Rempel2014}. From the simulation cube, we extracted a sequence of synthetic intensity maps at $\lambda=500$~nm and their corresponding horizontal velocity field at optical depths $\tau =[1, 0.1, 0.01]$, where the optical depth is defined at the continuum wavelength of 500~nm. These snapshots have a full FOV of $34'' \times 34''$ with a spatial resolution of $0.022''$ ($\sim$16 km), and the sequence has a 10-second cadence over a duration of one hour. Following \citet{AsensioRamos2017}, we normalized all input data to the median intensity of the quiet Sun and scaled the velocities to the interval $[0, 1]$. The training dataset was generated by degrading the simulated data to match the $0.054"$ sampling of the context images. This process included histogram matching to align the distributions of the simulated continuum intensity and the observed continuum intensity at $\lambda \approx 590~$nm.

The resulting training set consisted of 16\,000 samples, each containing a consecutive pair of continuum intensity images and the corresponding horizontal velocity fields, with a spatial size of $128 \times 128$ pixels ($6'' \times 6''$). The model was trained for 100 epochs using the Adam optimizer \citep{Kingma2014} with a learning rate of 0.001. To ensure a robust validation, we spatially partitioned the simulation into four equal quadrants, extracted training samples from the first three quadrants, and reserved a spatially independent validation set of 1600 samples from the fourth quadrant. Finally, the trained model was applied to the MiHI context images (continuum around 590~nm), binned down to the cadence of 10~s, to match the simulation. Examples of derived horizontal velocities are given in Fig.~\ref{fig:horiz_vel}. 

\section{Simulations and spectral synthesis}
\label{sec:simulations}

Although the high-resolution, fast-cadence, high-spectral fidelity MiHI data allow us to study various physical properties of the collapsing granule, our observational analysis has limits; mostly in the sense that it is impossible to unambiguously reconstruct the fully 3D physical state and the energy balance of the atmosphere during the observed collapse. Even depth-dependent, physics-regularized spectropolarimetric inversions \citep[e.g.][]{borrero_mhs_I} struggle to retrieve accurate physical parameters for such dynamic atmospheric events. To further investigate the physics of this specific event, we turn to a comprehensive, high-resolution, radiative-magnetohydrodynamic simulation of the lower solar atmosphere.

\subsection{CO5BOLD simulation}
\label{ssec:coboldsim}

The simulation was carried out using the CO5BOLD code \citep{2012JCoPh.231..919F}. The code solves the equations of ideal MHD for a fully compressible gas in a Cartesian box and is based on a conservative finite-volume scheme using an approximate Riemann solver \citep[Roe, HLLMHD;][]{2017MmSAI..88...37S}. It uses a general equation of state that adequately describes the solar plasma, with the chemical elemental abundances taken from the CIFIST project \citep{2009MmSAI..80..711L}. The radiative transfer step proceeds via a multi-group scheme using tabulated realistic opacities provided as a look-up table with twelve bins, using opacities from the MARCS model atmosphere package \citep{2008A&A...486..951G}. CO5BOLD has been extensively used to study small-scale events and waves in the solar atmosphere. The simulation run in this work was carried out with the latest version of the code, which uses hybrid MPI/OpenMP parallelization \citep{2022A&A...660A.115R,2024A&A...684A...7R}.

For this work, the 3D magnetoconvection model was obtained from a relaxed model of solar convection computed using CO5BOLD \citep{2018Calvo_thesis}. The model was interpolated to double its original resolution, and a homogeneous vertical field of 50~G magnetic flux density was embedded in the model at $t=5646$~s. The new computational domain spans $9.6\times9.6\times2.56$ Mm$^3$ in the two horizontal and one vertical direction, discretized on $1920\times1920\times512$ grid cells of equal size. The vertical extent of the box covers from $z=-1.24$~Mm below to 1.32~Mm above the mean Rosseland optical depth $\tau_{\text R} = 1$ surface.  A constant external gravity field with $g = 275~{\rm m~s^{-2}}$ acts along the $z$ direction.

The top boundary is open for fluid flow and outward radiation, while an exponential decrease of the mass density is imposed for the boundary cells outside the domain. The vertical component of the magnetic field is constant across the boundary, and the transverse component drops to zero at the boundary. The bottom boundary is set up in such a way that the in-flowing material carries a constant specific entropy of $1.775 \times 10^9~{\rm erg~g^{-1}~K^{-1}}$ resulting in a radiative flux corresponding to an effective temperature of $T_{\rm eff} \approx 5770$~K. The bottom boundary conditions for the magnetic field are the same as for the top boundary. Periodic boundary conditions are used for the sides. 

\begin{figure}[htbp]
    \centering
    \includegraphics[width=1.\columnwidth]{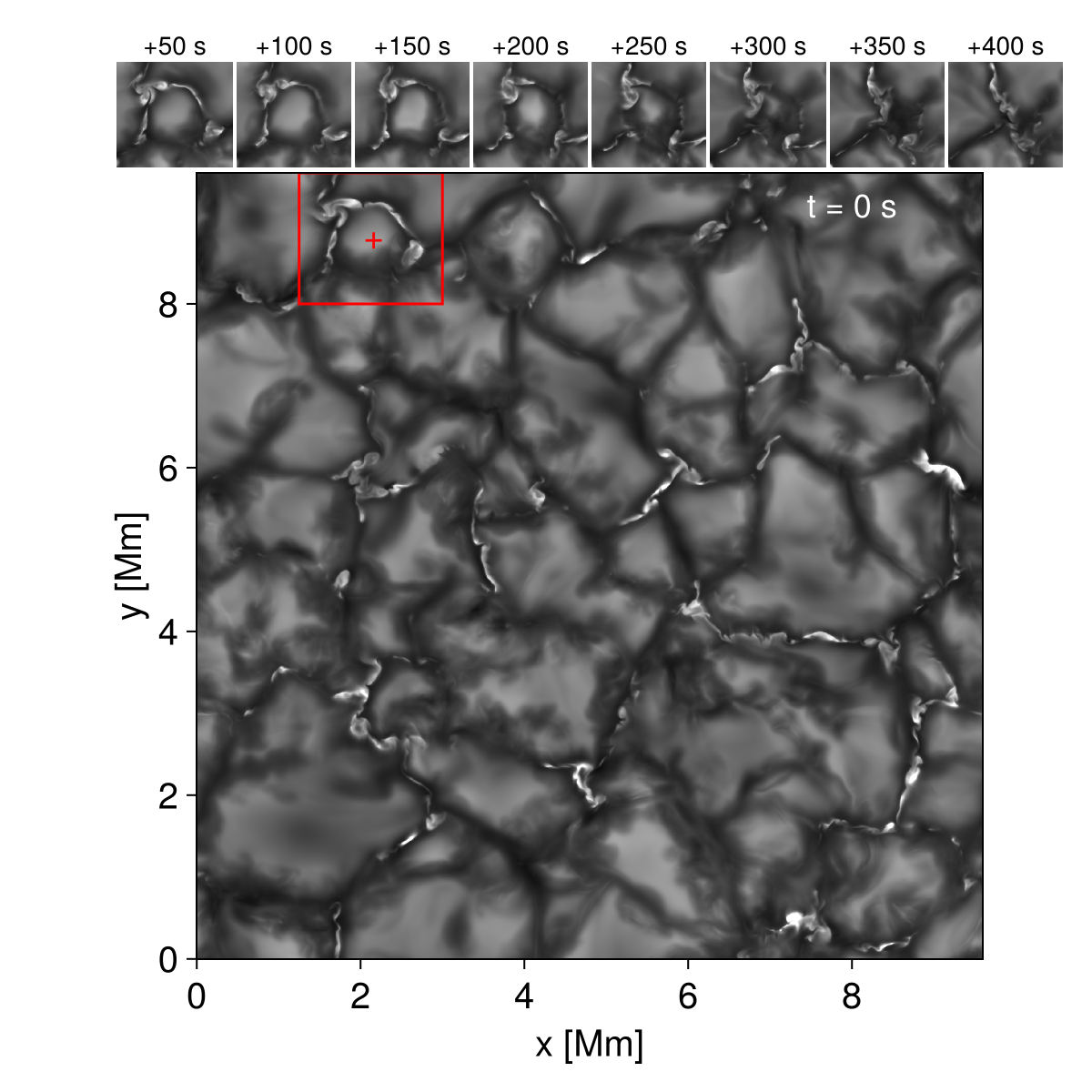}
    
    \caption{Emergent bolometric intensity in the full field of view in the CO5BOLD simulation at $t=0$~s. The top panel displays the sequence of snapshots at different instants (every 50~s) showing the granular collapse event in the region marked in red in the panel below. The red-cross hair marks the location where the temperature has a local maximum at $z=0$~km, which we define as the center of the granule.} 
    \label{fig:sim_gran_collapse}
\end{figure}

We focus on a region centered around the point $(x,y) = (2.1, 8.8)$~Mm (see Fig.~\ref{fig:sim_gran_collapse}). Here, a small granule collapses, starting at approximately $t=8494$~s. From now on, we refer to this time as $t=0$. The timescales and $t=0$ reference times are defined separately for the observation and the simulation, and the meaning of $t$ should be clear from the context. The collapse event lasts roughly 450~s. At the point $(x,y)$ marked by the red cross-hair the temperature has a local maximum in the small FOV at $z=0$~km\, defined as the layer where the mean $\tau_{R}=1$, at $t=0$~s and we use this location to define the center of the granule. We chose this event, out of the four granular collapses identified in the simulation, because of the regular shape of the granule. Note that we did not perform a detailed comparison between the observed and simulated collapses before the selection of the event. The size of the granule and the duration of the event are in agreement with the observed granular collapse, as can be seen from the comparison of Fig.\,\ref{fig:inv_evol} with Figs.~\ref{fig:sim_gran_collapse} and \ref{fig:horiz_vel}, and also in agreement with the statistical study of \citet{Mueller_2001_granules}. To follow the event more closely, we captured this region with a cadence of 1~s, between $z=-740$ to 760~km. The selected FOV covers the region from 1.25~Mm to 3~Mm in the $x$ and 8~Mm to 9.6~Mm in the $y$ direction, denoted with a red square in Fig.~\ref{fig:sim_gran_collapse}.

\subsection{Spectral synthesis}
\begin{figure}
    \centering
    \includegraphics[width=1\linewidth]{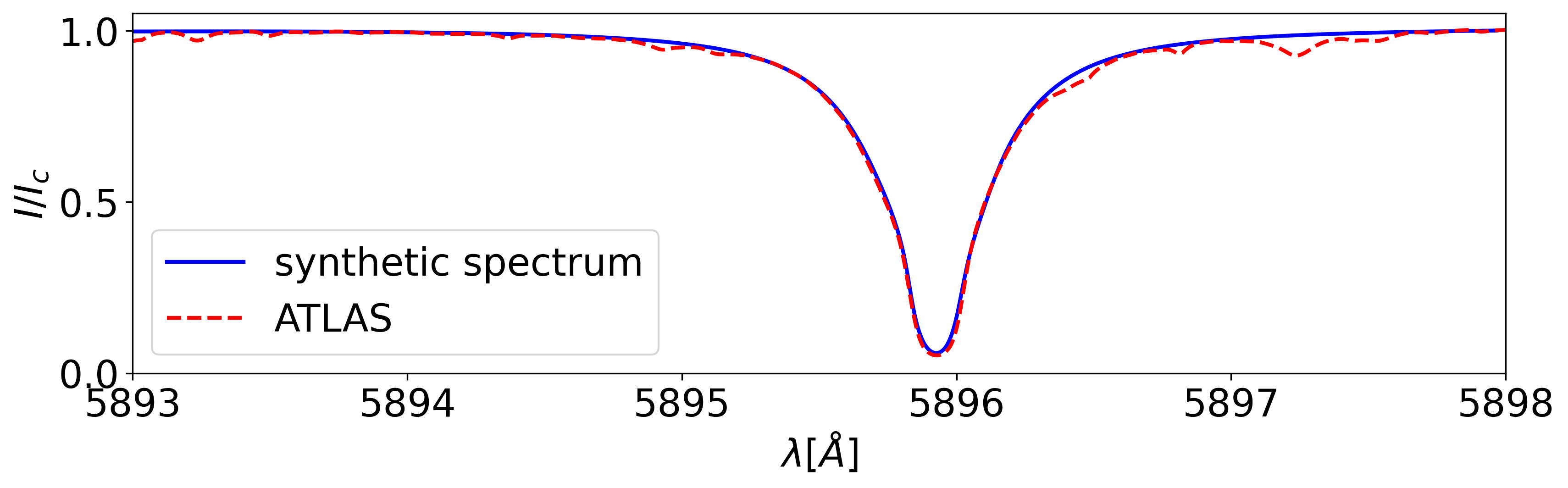}
    \caption{Synthetic Stokes $I$ \ion{Na}{i} D1 line averaged over the area occupied by the granule (see red box in Fig.\,\ref{fig:sim_gran_collapse})     compared to the solar atlas of \citet{Delbouille_1973_atlas}.}
    \label{fig:atlas_comparison}
\end{figure}

To calculate the Stokes spectra in the wavelength region observed by the MiHI from the simulated time series, we used the Lightweaver NLTE radiative transfer framework \citep{Lightweaver_Osborne_2021}. The atmospheric parameters used for the synthesis are temperature, gas pressure, line-of-sight velocity, and magnetic field vector, where each vertical column was treated as a semi-infinite one-dimensional atmosphere (the so-called 1.5D approximation). We calculated the emergent Stokes spectra of the \ion{Na}{i}\,D1 line in the spectral range of $5893-5898$~\AA, with $5~{\rm m\AA}$ sampling. We used a \ion{Na}{i} atom model with $12$ atomic levels, accounting for non-LTE excitation and ionization of sodium, assuming statistical equilibrium. The synthesis was performed for every second output step in the simulation, leading to a cadence of $2$~s, comparable to the MiHI observations. Fig.~\ref{fig:atlas_comparison} illustrates the excellent agreement between the mean synthetic \ion{Na}{i}\,D1 spectral line profile and the solar atlas by \citet{Delbouille_1973_atlas}.  A side-by-side comparison of the observed and simulated granular collapse is available as an \href{https://www.dropbox.com/scl/fi/vk0gy9moqrbbdbihlbn5s/collapse_obs_vs_sim.mp4?rlkey=hwqrs8hw5gv51j0fvo9t5c4x6&st=7wknnrdu&dl=00}{animation}. The associated movie is available online.  

\section{Comparison between simulation and observations}
\label{sec:comparison}

\subsection{Granulation properties}
We start by comparing the granulation properties between the simulation and the observations. In the simulation, the mean horizontal scales computed from the power spectra of the bolometric intensity over the full FOV and averaged over 400~s show maxima in the range of 1600 - 2000~km \citep[in agreement with e.g.,][]{Hirzberger_1997ApJ_granulesize}. At smaller scales, the spectra follow a power law down to the dissipation scales in the simulation (around 45~km), after which we see a sharp decline in the size distribution until the Nyquist limit of 10~km. The upper limit in the granular size is essentially set by the effective temperature and surface gravity of the star \citep[][]{RAst_1995_granules,1998ApJ...499..914S,   Nordlund_2009_review}, which is also confirmed by the simulations presented in, e.g., \citet{Beeck_2013_conv} and \citet{Trampedach_2013_grid}. The lower limit in the feature size distribution we have here is of numerical origin, and well below typical granular sizes found in the simulations.

The specific granule we focus on in this study has a diameter of approximately 750~km before the collapse, which places it well within the inertial range of the size cascade. A recent analysis of observed granulation, by \citet{2021ApJ...923..133L}, shows that the granules fall into three length ranges, separated by two critical sizes of 265 and 1420~km, where the lower threshold marks a turbulence-dominated granular evolution regime and the upper threshold a convection-dominated one. The central regime to which the granule considered here belongs suggests that both convection and turbulence are relevant in its evolution.

For the observations, we estimate the spatial scales from the power spectra of the maps with a FOV of $44''\times44''$ (approximately $32 \times 32~{\rm Mm}^2$), provided by the context imager, which has a slightly finer spatial sampling (around $0.054''$) compared to the spectral data. The granulation pattern exhibits a trend in spatial scales similar to the simulation down to the resolution limit of 78~km, with a peak in the range larger than 1600~km. The specific granule that we focus on in this study has an effective diameter of $700$~km (corresponding to an area of $3.8 \times 10^5~{\rm km}^2$) before the collapse, which aligns with the pre-collapse dimensions of the simulated granule.

\subsection{The collapse event}
\label{ssec:collapse_event}

Granules typically evolve by expanding horizontally near the surface -- the central region eventually cools more rapidly than can be compensated for by the hot upflowing material. This leads to fragmentation of the original granule into smaller ones. In some cases, smaller granules collapse rather than expanding again. This may happen either due to the lower enthalpy flux from below within the granule or due to the granule being squeezed by the neighboring granules. In the former case, the critical value for the vertical velocities is found to be around $2~{\rm km~s^{-1}}$ \citep{1985SoPh..100..209N}. 

\begin{figure}[htbp]
    \centering
    \includegraphics[width=1.\columnwidth]{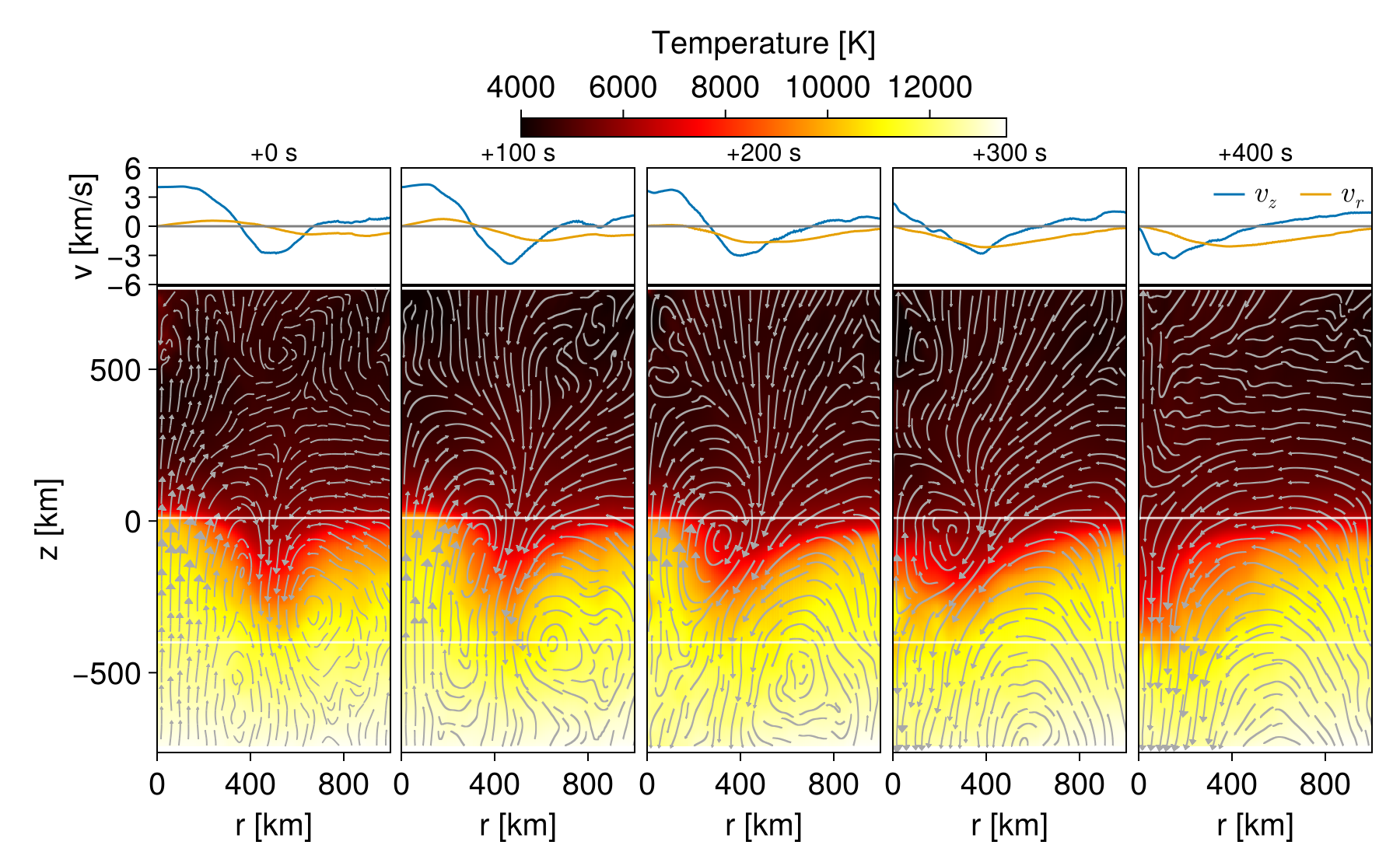}
    \caption{Time sequence of the azimuthally averaged temperature in the $(r,z)$ plane showing the granular collapse in the simulation. Here $r$ is the radial distance from the center of the granule located at $(x,y) = (2.16, 8.78)$~Mm, represented by the red crosshair in Fig.~\ref{fig:sim_gran_collapse}. The top panels show the azimuthal averages of the radial ($v_r$; in orange) and vertical component ($v_z$; in blue) of the velocity. Times are given with respect to the start of the collapse. Streamlines track the plasma velocity in the $(r,z)$ plane.}
    \label{fig:sim_granule_radial_temperature}
\end{figure}

\begin{figure}
    \centering
    \includegraphics[width=1\linewidth]{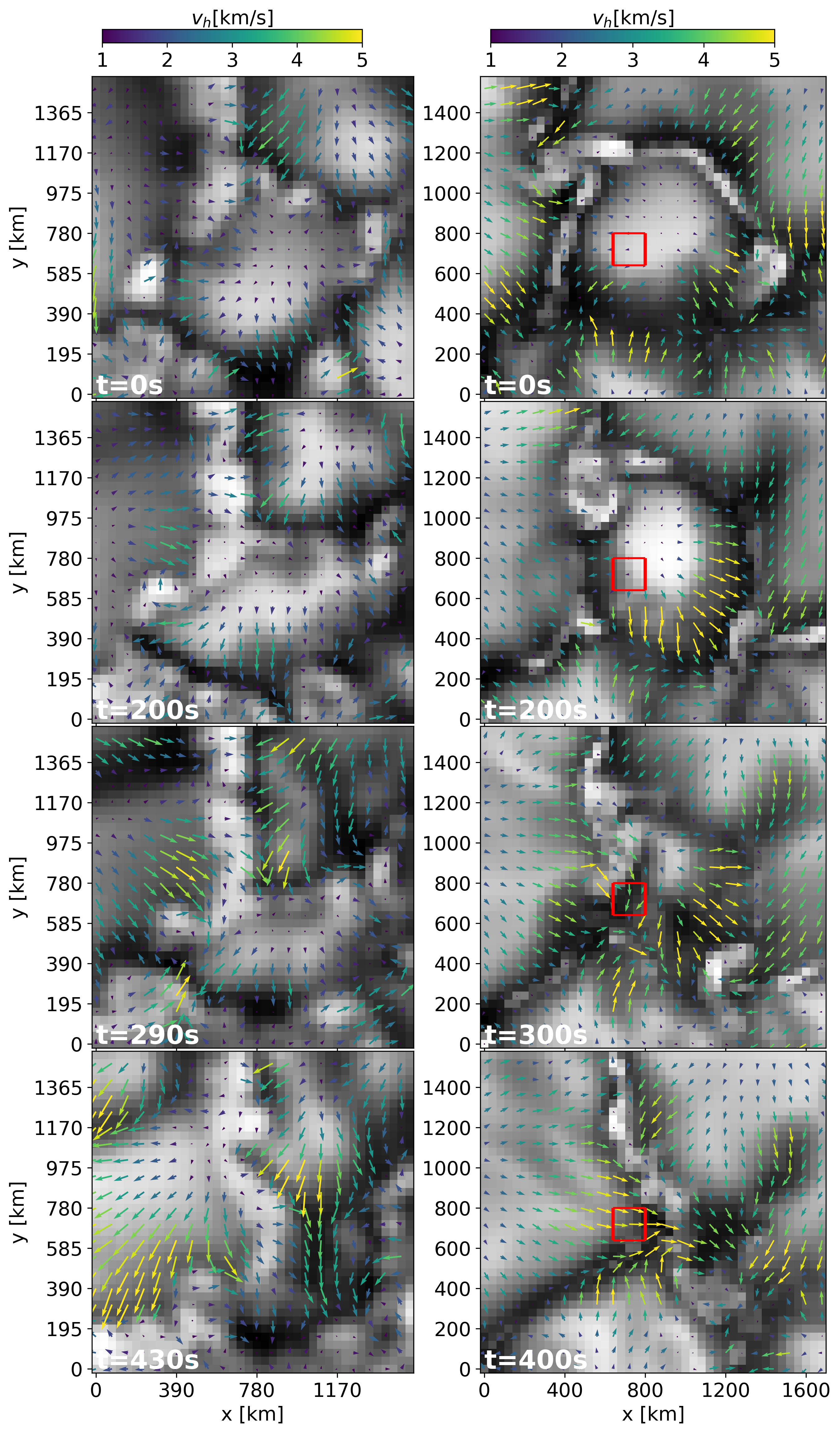}
    \caption{Time evolution of the continuum intensity and the horizontal component velocity field. The left column shows the observations, and the right column the simulation binned to match the resolution of the observations. The arrows indicate the direction, and the color indicates the magnitude of the horizontal velocities. Time is given with respect to the start of the collapse. The red square in the right column marks the ROI.}
    \label{fig:horiz_vel}
\end{figure}

Figure~\ref{fig:sim_granule_radial_temperature} summarizes the time evolution of azimuthally-averaged temperature and the flow field in the considered granule during the collapse in the simulation. At the start, there is an initial upflow in the granule of around 4\,km~s$^{-1}$, which is above the critical ascent velocity required to sustain a granule. At the same time, there exists a strong radially inward flow from the surrounding granules of around 2-3\,km~s$^{-1}$. Fig~\ref{fig:horiz_vel} shows the intensity and the horizontal component of the flow velocity, inferred from the observations and retrieved from the simulation. The simulated intensity and velocity are binned to match the resolution of the observations. The flow velocity reaches 5\,km~s$^{-1}$ towards the end of the collapse. The horizontal flow from the surrounding regions squeezes the granule, heating it and thus increasing radiative losses, eventually leading to its collapse, with cool downflows of around 3\,km~s$^{-1}$ replacing the initial granular upflows. We also note that the granule collapses asymmetrically with respect to its center, as it gets pushed by the larger nearby granules, which exhibit accelerating horizontal flows towards the collapsing one. This evolution is evident from Fig.~\ref{fig:horiz_vel}, which shows the temporal evolution of the continuum intensity and the corresponding horizontal velocities for both the observations and the simulation.  The timescales of the two events are comparable, and both show the presence of bright magnetic flux concentrations around the granule, which eventually coalesce into a single one (also see columns 3 and 4 in Fig.~\ref{fig:inv_evol}).

The left column of Fig.~\ref{fig:horiz_vel} shows the horizontal velocities inferred by the DeepVel model from the time series of the MiHI context images. Overall, the magnitude, direction, and evolution of these horizontal velocities are comparable to the simulated ones. This demonstrates the ability of the trained model to identify the particular intensity variations and to infer a coherent horizontal velocity field of a collapsing granule, considering that the DeepVel model was trained on a completely independent numerical simulation, carried out with a different code and with a different spatial resolution. Finally, the evolution of the vertical velocities inferred from the observations (second row of Fig.~\ref{fig:inv_evol}) shows that the original granular upflow is eventually replaced by a downflow. 

\begin{figure}
    \centering
    \includegraphics[width=\linewidth]{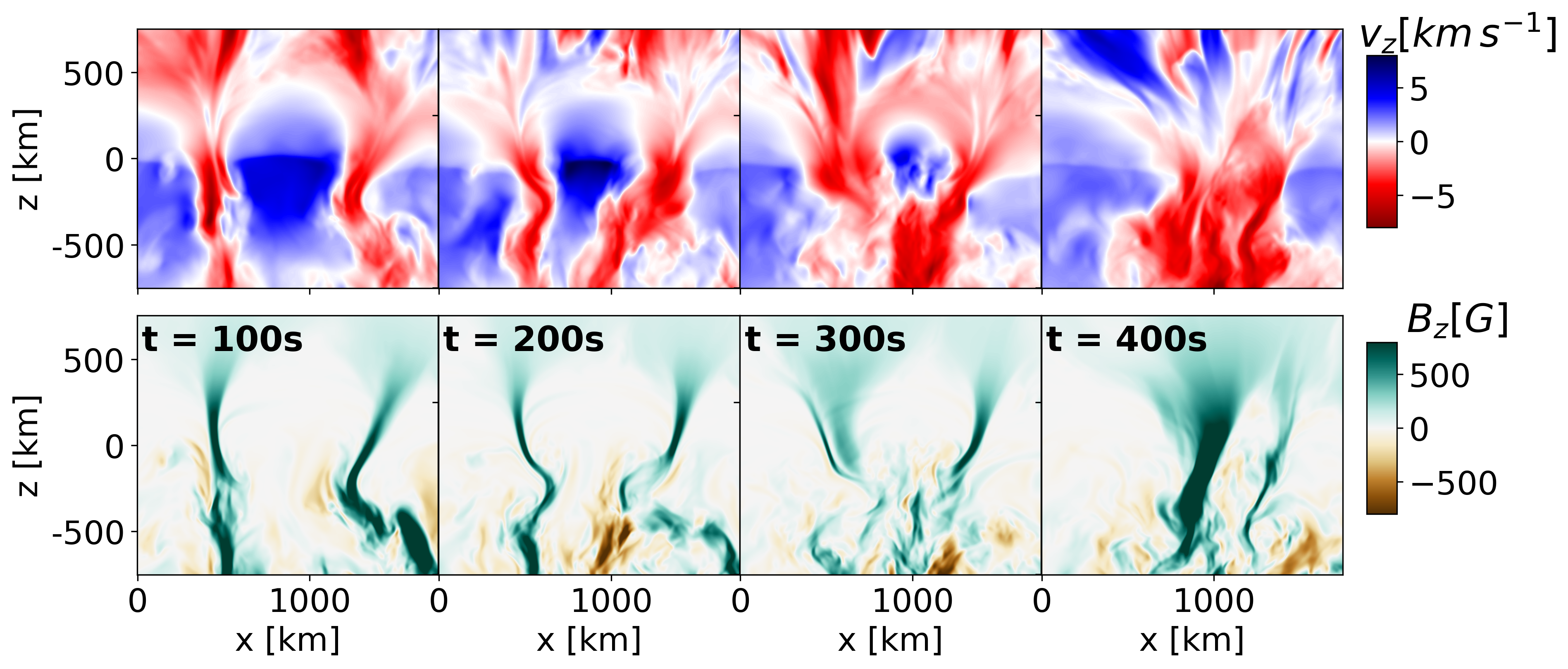}
    \caption{Underlying vertical velocity and magnetic field during the simulated collapse. The quantities are averaged in y-direction  over 150km in the ROI.}
    \label{fig:underlyingdynamics}
\end{figure}

The underlying dynamics of the simulated collapse process are shown in Fig.~\ref{fig:underlyingdynamics}. In the vertical velocities we see strong downward flows below the collapsing granule that increase with time. In agreement with this, \cite{Skartlien2000collapse} report that collapsing granules are located above subsurface downflows. In addition, we observe a concentration of the magnetic field and an increase in its strength, during the later stages of the collapse.

\begin{figure}
    \centering
    \includegraphics[width=1\linewidth]{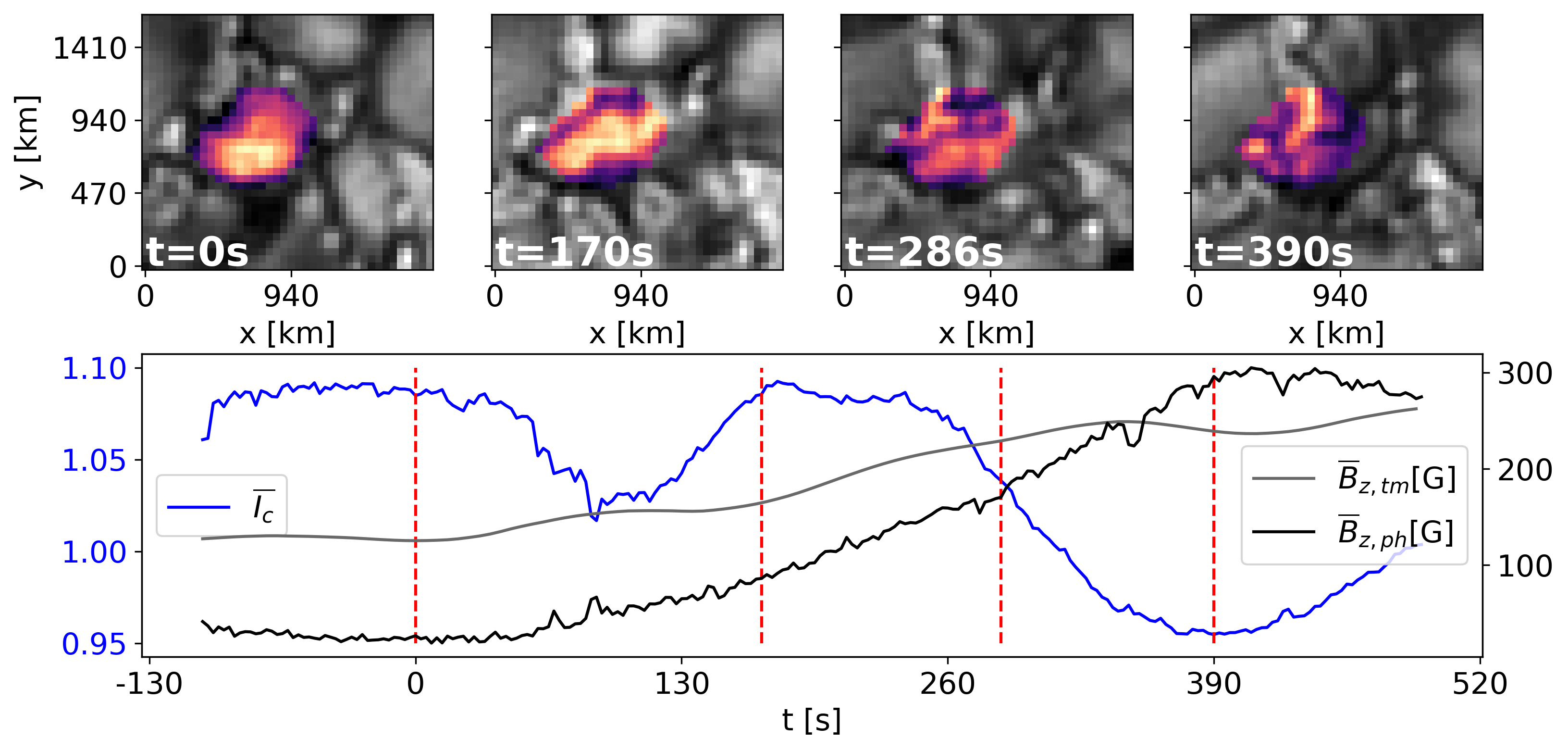}
    \includegraphics[width=1\linewidth]{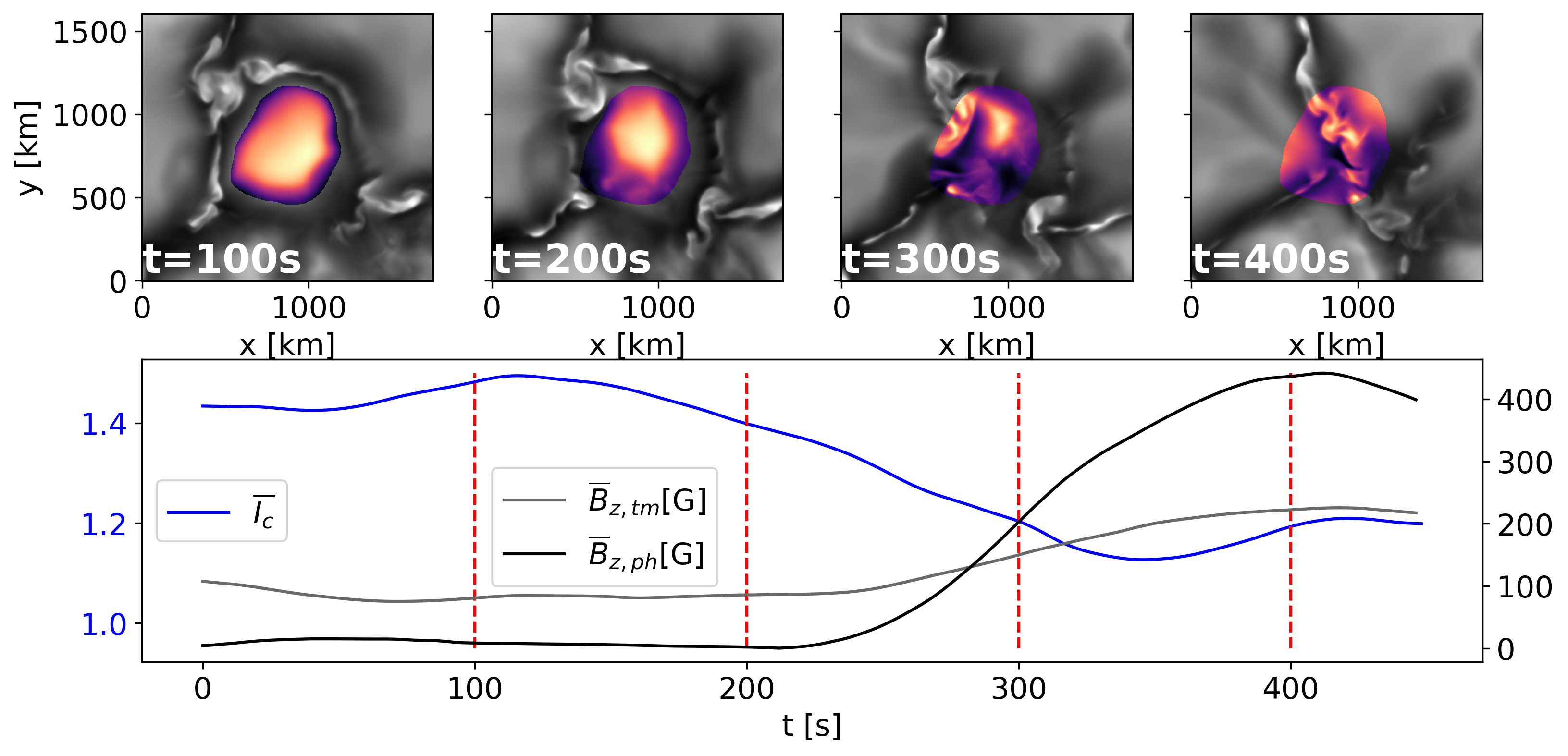}
    \caption{Evolution of the mean physical parameters within the area occupied by the collapsing granule. The top row shows the evolution of the observed continuum intensity with the mask overlaid in color. The second row shows the time evolution of the continuum intensity (left axis) and the vertical magnetic field in the photosphere and in the temperature minimum (right axis), averaged over the masked surface. The dashed red lines mark the time steps that correspond to the images in the first row. The bottom two rows show the corresponding plots for the simulation.}
    \label{fig:mask}
\end{figure}

To track changes in brightness and magnetic field within the area occupied by the collapsing granule, we defined a spatial mask based on the granule's shape at the onset of the collapse. This mask is defined by a continuum intensity contour threshold of $0.95~I_{c}$. Figure\,\ref{fig:mask} shows the evolution of the mean intensity $\overline{I_c}$ and magnetic field within the mask, for the observation (top) and the simulation (bottom). From the decreasing mean intensity and increase of the mean magnetic fields in the photosphere and the temperature minimum, it is evident that the collapse leads to a concentration of magnetic field in both cases. In the observation, $\overline{I_c}$ decreases from the value of $\approx$1.08 at the start of the collapse, to roughly $1.02$ one minute after, increases again over the next $\approx$2~minutes, and then decreases down to 0.95 by the end of the collapse. In the simulation, the intensity increases right at the start of the collapse. This increase in brightness at the beginning of the collapse (as seen in Fig.~\ref{fig:mask} for the first 115 seconds in the simulation) has already been reported by \citep{Skartlien2000collapse} and is explained by an initial enhancement of the upward enthalpy flux. 

The mean vertical magnetic field strength in the photosphere within the mask monotonically increases in time from approximately  $\overline{B}_{z,ph}=0$~G to 300~G, both in the simulation and the observation. The corresponding magnetic field at the temperature minimum, $\overline{B}_{z,tm}$, increases from approximately 100~G to 250~G in the observations and behaves qualitatively similarly in the simulations. The evolution of the mean vertical field in the photosphere and the temperature minimum differ, because the expansion of the magnetic field with height leads to different spatial distributions of the field between the photosphere and the temperature minimum (as visible from Fig.~\ref{fig:inv_evol}). Namely, at the start of the collapse, the masked area contains no magnetic flux concentrations in the photosphere, but, due to the magnetic flux tube expansion with height, the field from the intergranular lanes expands into the masked area towards the temperature minimum. Similarly, at the end of the collapse, part of the temperature minimum field remains outside of the masked area.

\subsection{Identification of a wave-like event}

\begin{figure}
    \centering
    \includegraphics[width=1\linewidth]{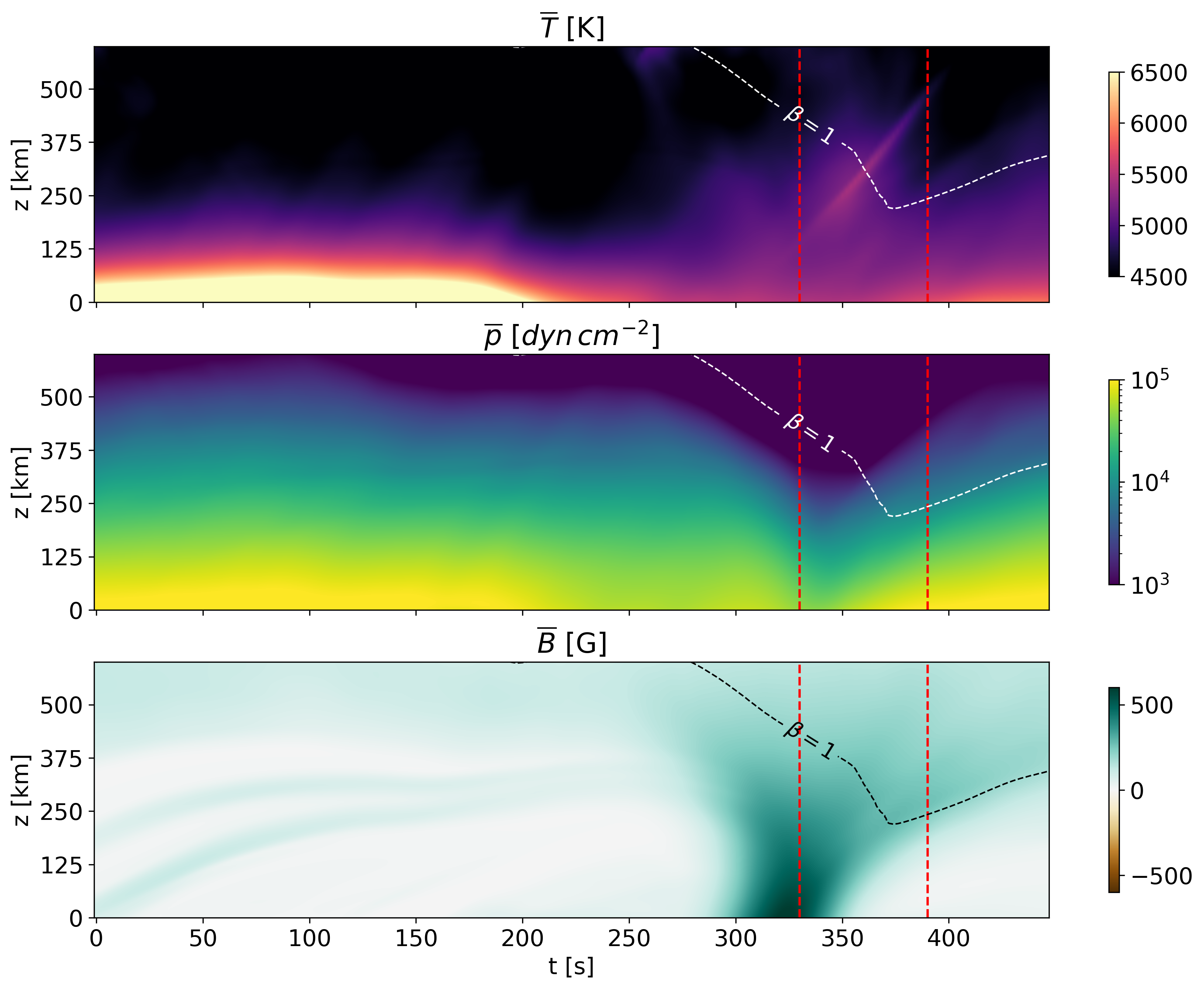}
    \caption{Time-height diagram of the mean temperature (top), gas pressure (middle), and the magnetic field strength (bottom) in the simulation, averaged over the 150$\times$150 ${\rm km}^2$ ROI, denoted with a red square in Fig.~\ref{fig:horiz_vel}. The vertical dashed red lines mark the time interval ($t=330$--$390$~s) during which the upward-propagating wave perturbation is detected. The contour line indicates plasma $\beta$=1.}
    \label{fig:wave_detection}
\end{figure}

Towards the end of the simulated collapse ($t\approx300$~s), magnetic flux concentrations around the granule get dragged together, causing a decrease in the density and, consequently, gas pressure (also shown in Fig.~\ref{fig:underlyingdynamics}). We detected an accompanying perturbation that propagates upward and exhibits a temperature enhancement of $500-1000$~K relative to the background. 

The accumulation of the magnetic field in the newly formed downflow region is also seen in the observations (Fig.~\ref{fig:inv_evol}). This strong magnetic flux concentration seems to play an important role, likely channeling the wave when propagating toward the upper layers \citep[see also, ][]{Jess2023_wave_review}. The exact spatial origin of the wave is hard to determine, as also pointed out by \citet{Kitiashvili2019_mag_wave_exc_ApJ...872...34K}.

To better understand the atmospheric response to the changes in the near-surface region due to the collapse, we analyzed the evolution of thermodynamic quantities over a $150 \times 150~{\rm km}^2$ area above the simulated collapsing granule, denoted by the red square in Fig.~\ref{fig:horiz_vel}. Hereafter, we refer to this as the region of interest (ROI). In Fig.\,\ref{fig:wave_detection} we display the time-height diagrams corresponding to the collapse in the ROI. The perturbation is clearly detectable in temperature, taking place at the times between the dashed red lines. Based on this, we define the wave event to start at $t=330$~s and to end at $t=390$~s.

\begin{figure}
    \centering
    \includegraphics[width=0.5\textwidth]{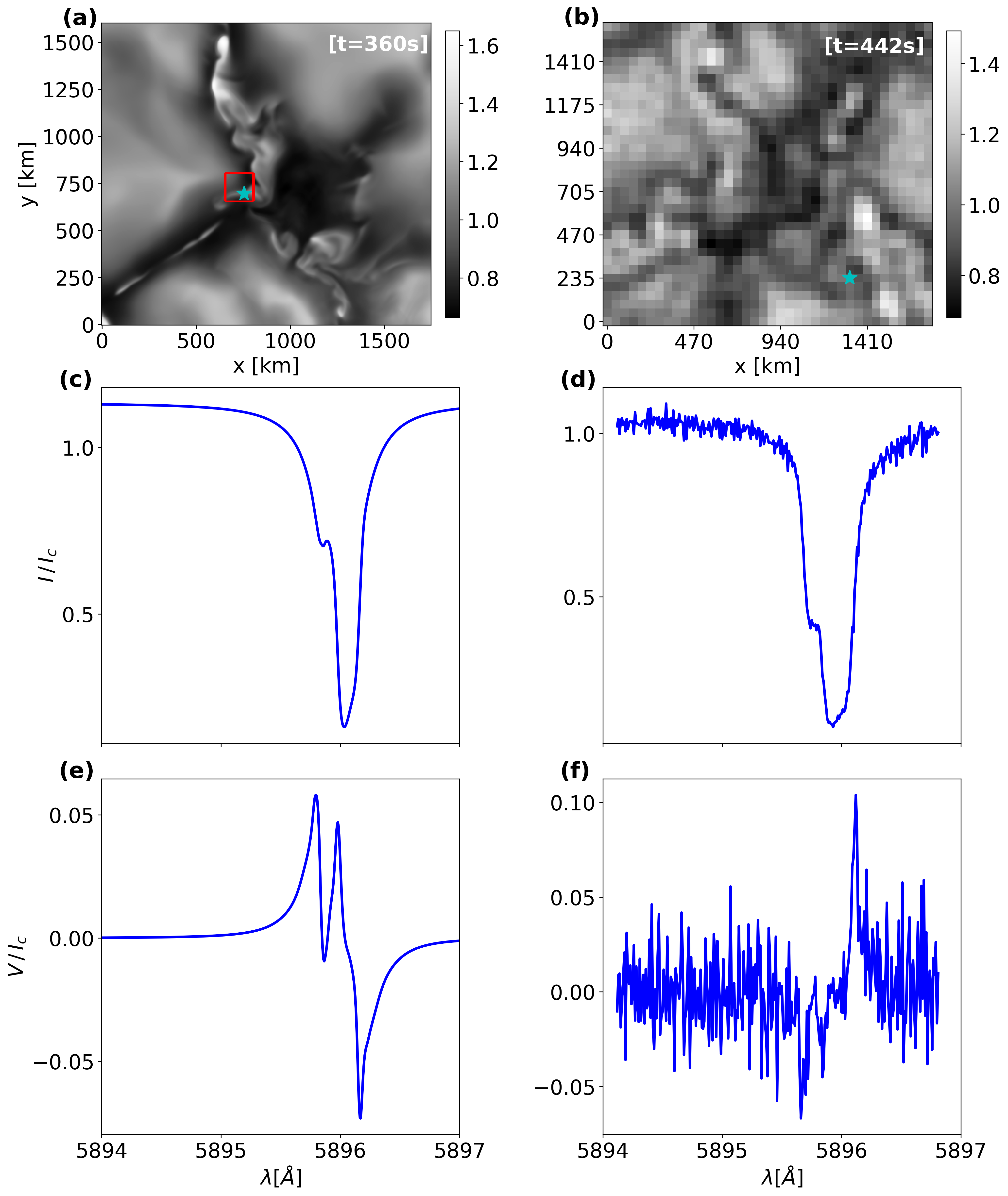}
    \caption{Spectral features found in the synthetic spectra (left column) and the observations (right column). Panels a) and b): continuum intensity map at the moment the \ion{Na}{i} blue wing emission is detected. The positions where the signatures are found are marked by cyan asterisks. Panels c) and d): Stokes $I$ profiles at the marked locations. Panels e) and f): Stokes $V$ profiles at the marked locations. The observed Stokes $V$ is integrated over 3 time steps ($\approx$8\,s). Red square marks the ROI.}
    \label{fig:wave_sign}
\end{figure}

To further strengthen the relevance of the simulated collapsing granule, we compared the synthetic \ion{Na}{i}~D1 spectra to observations. The synthetic spectra emerging from the ROI show an emission peak in the blue wing of the \ion{Na}{i}~D1 line that corresponds to the propagating feature in the simulation (Fig.~\ref{fig:wave_sign}c and e). The observed spectra exhibit the same blue-shifted emission feature during a similar stage of the collapse (location marked by the cyan asterisk in Fig.~\ref{fig:wave_sign}b). This spectral signature appears at $t=360$\,s in the simulation and $t=442$\,s in the observation. The corresponding Stokes $V$ line profiles are also remarkably similar, except for the different sign due to the opposite field polarities (Fig.~\ref{fig:wave_sign}e and f). 

\subsection{Wave energy flux and propagation}
\label{ssec:flux_and_prop}

We estimate the energy flux of the traveling perturbation in the simulation following the discussion in \cite{2003ApJ...599..626B} and use:
\begin{equation}
    \vec{F}_{\rm wave}=\Delta p\, \vec{u}
\end{equation}
to calculate the acoustic energy flux of the wave in the $z$-direction, where $\vec{u}$ is the fluid velocity and $\Delta p$ is the perturbation of gas pressure. We define $\Delta p$ as the difference between the pressure at time $t$ and time $t=0$.

\begin{figure}
    \centering
    \includegraphics[width=1\linewidth]{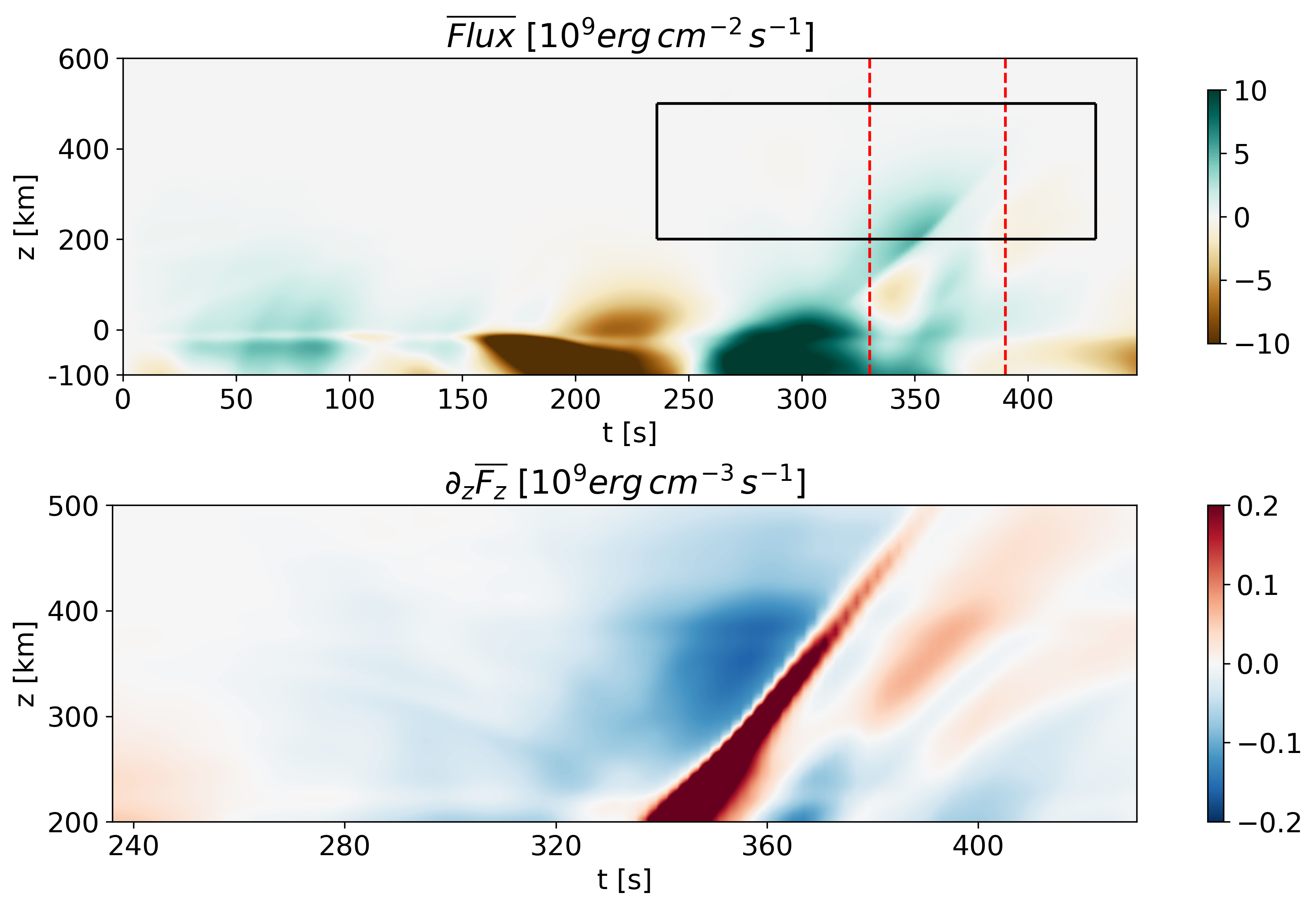}
    \caption{Top: Height-time diagram of the acoustic flux averaged over the ROI in the simulation. 
    The dashed red lines again mark the time during which we detect the wave perturbation. The origin of the $z$ axis is defined with respect to the base of the photosphere ($\log\tau=0$). Bottom: The vertical component of the flux divergence in the area indicated by the black box in the height-time diagram at the top.}
    \label{fig:flux_whole_sim}
\end{figure}

Figure~\ref{fig:flux_whole_sim} shows the evolution of the height-dependent acoustic energy flux throughout the whole collapse event, averaged over the ROI. Halfway through the collapse ($t \approx 200--250$~s), there is a strong downward flux, followed by a strong upward flux, which is, finally, followed by the wave excitation. This behavior has also been reported by \citet{Skartlien2000collapse}. The evolution of the flux in Fig.~\ref{fig:flux_whole_sim} suggests that the wave originates slightly lower than indicated by the temperature perturbation in Fig.~\ref{fig:wave_detection}.

\begin{figure*}
    \centering
    \includegraphics[width=1\linewidth]{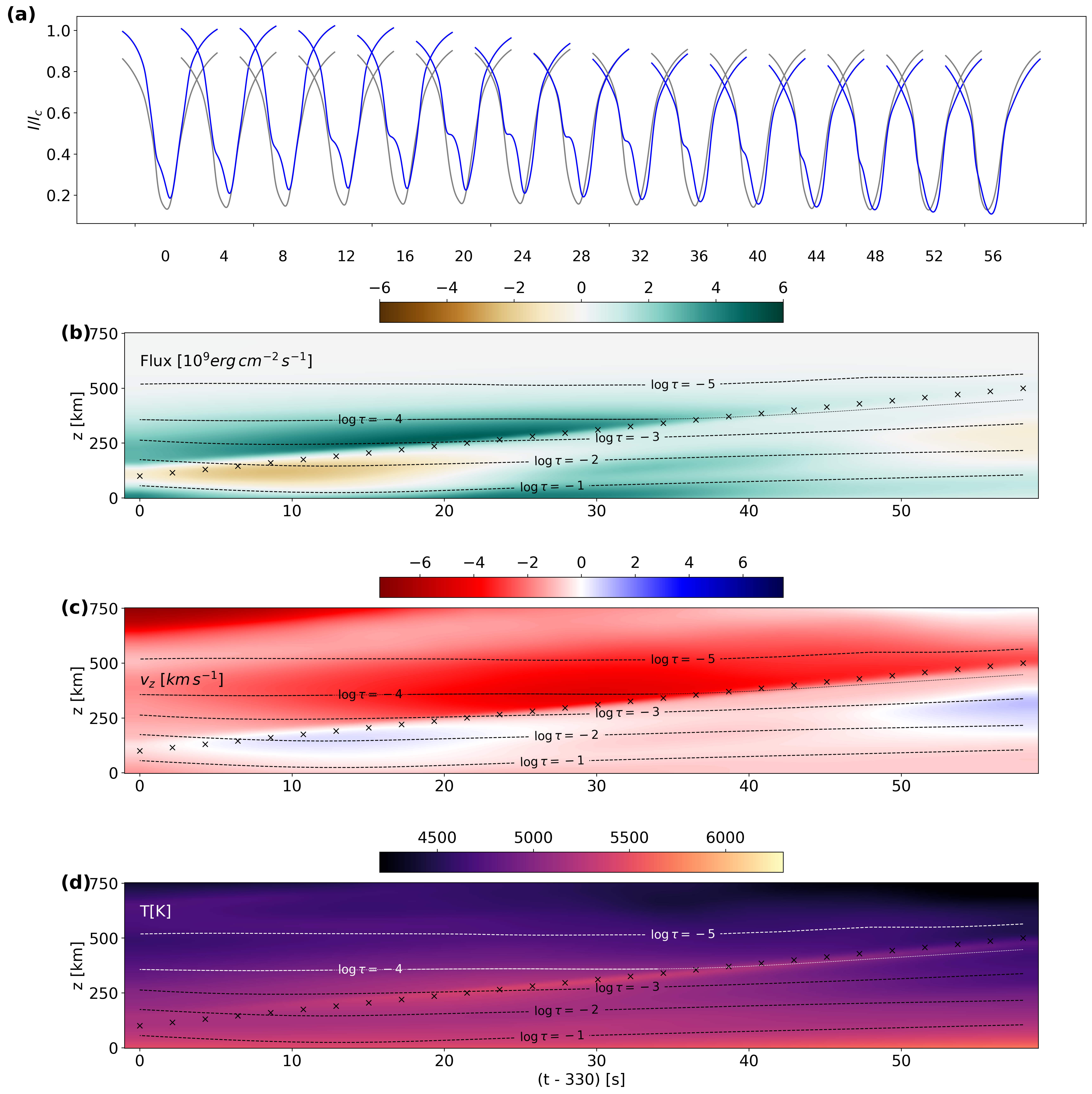}
    \caption{Simulated quantities during the wave propagation (t=330s-390s): Panel a): Time evolution of the spectra averaged over ROI (blue lines), and the mean spectra over the full FOV (gray lines) during $t=$330-390~s, indicated by dashed red lines in Fig.\,\ref{fig:flux_whole_sim}. Panel b): Time-height diagram of the acoustic energy flux, averaged over the ROI. Panel c): Time-height diagram of the line-of-sight velocity averaged over the ROI (red/negative: downward flow, blue/positive: upward flow). Panel d): Time-height diagram of the temperature averaged over the ROI. The dashed lines in panels b), c) and d) show the layers of constant optical depths, and the crosses show the trajectory of a test particle traveling at local sound speed.}
    \label{fig:flux_mean_over_time}
\end{figure*}

In Fig.\,\ref{fig:flux_mean_over_time} we show the spectra of the \ion{Na}{i}~D1 line alongside the evolution of the flux and temperature stratification with height, averaged over the ROI during the 60~s period that corresponds to the wave propagation. The enhanced emission in the blue wing of the \ion{Na}{i}~D1 line corresponds to a positive energy flux in the higher layers of the solar photosphere. This energy flux is correlated to the positive temperature perturbation that travels upward and is accompanied by the blue-shifted emission peak in the spectra. The crosses in Fig.\,\ref{fig:flux_mean_over_time} indicate a trajectory of a test particle traveling at the local sound speed. The perturbation in energy flux and temperature follows this trajectory (starting from $t\approx15$~s). This indicates that this traveling perturbation is essentially an acoustic wave. The local sound speed in the ROI is in the range of 7 to 8 km\,s$^{-1}$.

From Fig.\,\ref{fig:flux_mean_over_time}, we estimated the instantaneous acoustic flux transported by the wave to be $\approx\! 10^9~\fluxunit$ in the ROI. This is nominally a very high flux when compared to recent high-resolution studies. For instance, \cite{Molnar2023wave} estimated the wave energy flux at the heights probed by the \ion{Na}{i}\,D1 to be around $10^5-10^6~\fluxunit$ in a quiet sun region and $10^4-10^5~\fluxunit$ in a plage region. However, their study considers statistical properties of a longer time series and a larger field of view, while our value represents a highly localized, peak flux of an isolated acoustic pulse. We also see that the bulk of the energy is deposited around the $\log\tau=-4$ layer (i.e., the temperature minimum). This can be seen in the bottom panel of Fig.~\ref{fig:flux_whole_sim}. We observe a strong negative flux divergence in that region during the wave propagation, which indicates energy is deposited there.

\begin{figure}
    \centering
    \includegraphics[width=1\linewidth]{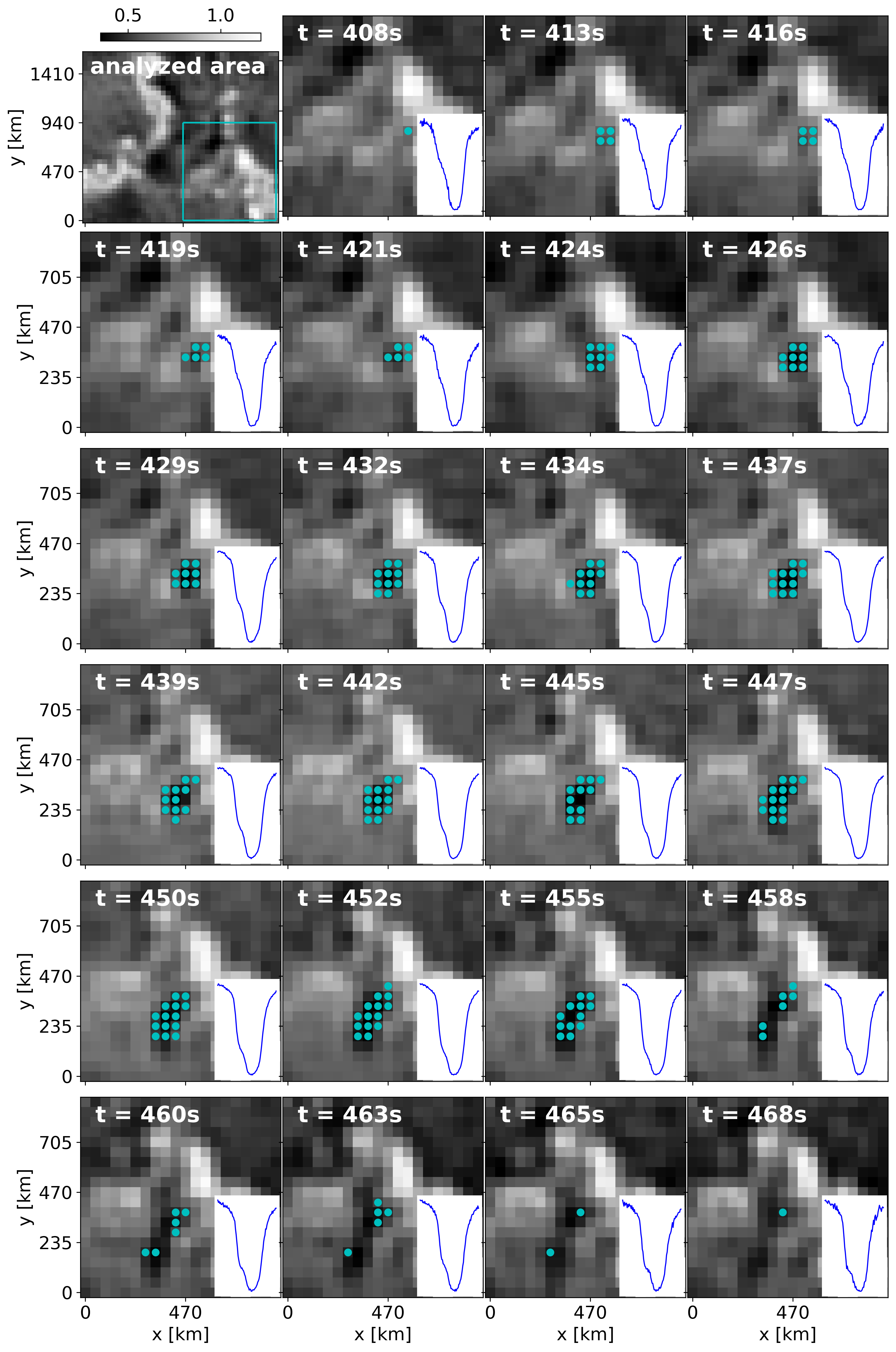}
    \caption{MiHI observations: Pixels marked with blue dots show the wave signatures similar to the one shown in Fig.\,\ref{fig:wave_sign} during the observed time series. The mean spectral line shapes of the \ion{Na}{i}\,D1 line in the indicated pixels are shown in the boxes in the lower right of each image.} 
    \label{fig:wave_duration}
\end{figure}

To further compare the simulation with the observations, we identify the specific pixels in the MiHI observations that exhibit the line wing emission that carries the wave signature. The evolution of the spatial distribution of these pixels is shown in Fig.\,\ref{fig:wave_duration}. In the observations, the first signature appears around $t=400$~s after the start of the collapse. The area occupied by the pixels exhibiting the wave signature increases from $t=406$~s from the start of the collapse until $t\approx450~$s, and then goes down again until the complete disappearance of the pixels with blue-wing emission, around $t=468~$s. Towards the end, the area occupied by the wave signatures seems to split into two sub-areas. The occurrence of these signatures might indicate that the wave is excited, takes $\approx60$~s to travel through the height range where the \ion{Na}{i}\,D1 line is sensitive, which is comparable to what we detect in the simulations. The detected blue-wing emission can be used as an indicator of such high-acoustic-flux events, and enable statistical characterization from observations with a larger field of view.

\section{Discussion and conclusions}
\label{sec:conclusions}

In this paper, we report on the analysis of a granular collapse event identified in observations performed with the MiHI \citep{vanNoort2022A&AMihiInstrument} instrument mounted at the Swedish 1-m Solar Telescope, alongside a similar event we identified in a high-resolution rMHD simulation of the solar photosphere performed using the CO5BOLD code \citep{2012JCoPh.231..919F}. Although this simulation was not explicitly set up, in any way, to reproduce this observed event, we found an excellent agreement between the granule sizes, duration of the collapse, and the evolution of the continuum intensity, magnetic, and velocity field. Even more remarkable is the excitation of a wave pulse toward the end of the collapse event. This perturbation travels upwards and leaves a clear imprint in the blue wing of the \ion{Na}{i}~D1 line, both in the observed spectra and in the synthetic observations produced from the simulations using non-LTE modeling.

Convective motions near the surface excite waves that can propagate to higher layers and deposit their energy via various dissipative processes, including shocks. Granular buffeting of the flux tubes embedded in intergranular lanes is a typical source of waves that has been studied extensively \citep{1981A&A....98..155S, 1998ApJ...495..468S_steiner_2d, 1999ApJ...519..899H_hasan_2d, 2012ApJ...755...18V_vigeesh_3d, 2014A&A...569A.102S_Stangalini_observations}, among the various other sources of waves in the lower photosphere \citep[see reviews;][]{Mathioudakis_2013_review, Jess2023_wave_review}. 
Acoustic emission is largely associated with the intergranular lanes but also found in regions where rapidly evolving small granules are present. 
Rapid changes in the granular structure are thought to be the main source of localized wave emissions. Earlier observational and numerical studies have shown that the main contribution comes from downward plumes that are initiated by rapid radiative cooling or are a part of collapsing granules \citep{Rimmele1995_wave_excitation,Rast_1999_thermal_plumes,Skartlien2000collapse,BelloG_2010_Sunrise_Acoustic, Kitiashvili2019_mag_wave_exc_ApJ...872...34K}. Interaction of vortices formed in the intergranular lanes is also thought to be a source of acoustic waves \citep{Kitiashvili2011_vortices_exc_acoutic}.

In this work, we found that a collapsing granule with a strong horizontal inflow from the surrounding region is followed by an impulsive, upward-traveling acoustic wave. 

The novel finding of this work is the identification of a specific wave event and its source through the unambiguous spectral signature, detectable thanks to the very high spectral and temporal resolution of the MiHI. This study complements recent efforts of \citet{2024ApJ...971L...1B} who detected signatures of an individual acoustic event in the quiet Sun, using continuum filtergram images from the DKIST solar telescope \citep{Rimmele2020dkist}. Recently, \citet{2026arXiv260207137M} studied the feasibility of detecting these quiet Sun events using an instrument akin to the Visible Tunable Filter \citep[VTF;][]{2025NatAs...9.1098K}, and discussed why such events manifest through the blue-wing emission in photospheric lines. While their results indicate that individual small-scale quiet Sun acoustic events are on the limit of detection of a 4-m telescope, we clearly identified and characterized a similar event at a lower resolution (50~km, the diffraction limit of a 1-m telescope). We hypothesize this detection may be related to the specific magnetic environment of the acoustic event that we have studied, and also to the fact that high spectral fidelity critically preserves the thermodynamic information content encoded in the line profile \citep{2025A&A...693A.272D}, allowing us to isolate the blue-wing enhancement.

The analysis of the wave excitation and propagation in the CO5BOLD simulation allowed us to quantify the acoustic energy flux carried by this wave. We found a value of the order of $10^9\fluxunit$, which is a factor of $10-10^3$ larger than what is reported in the statistical high-resolution studies \citep[e.g.,][]{Molnar2023wave, Joao2024wave}. For an adequate comparison with these works, an assessment of the temporal and spatial frequency of these events is necessary. According to \citet{Mueller_2001_granules}, 46.6 \% of all granules disappear by collapsing and their lifetime is $9.2\pm5.4$~min. \citet{Hirzberger_1997ApJ_granulesize} find 15.9\% disappearing by this mode with a mean lifetime of 6.03~min. Our analysis shows that the flux is concentrated in the ROI of linear size $l =150$~km, while the mean size of a collapsing granule is roughly $d\approx650$~km. Furthermore, this flux lasts only a short time of approximately 20~s. An estimate of the mean acoustic flux from these events is then:
\begin{equation}
    F_{\rm mean} = f_{\rm c} \times f_{\rm gr} \times \frac{l^2}{d^2} \times \frac{t_{\rm wave}}{t_{\rm lifetime}} \times f_{\rm wave} \times F_{\rm wave}
\end{equation}
where $f_{\rm c}$ is the fraction of the collapsing granules, $f_{\rm gr}$ is the fraction of the quiet Sun occupied by granules, $t_{\rm wave}$ is the duration of the wave event, and $t_{\rm lifetime}$ is the typical life time of a granule that disappears by collapsing. Finally, $f_{\rm wave}$ is the fraction of granules that excite wave events similar to the considered one, and $F_{\rm wave}$ is the acoustic flux carried by one such event.

Taking a conservative estimate for $f_{\rm c} = 0.2$, $f_{\rm gr}= 0.5$, and assuming that $t_{\rm wave} = 20$~s, $t_{\rm lifetime}=360$~s and $F_{\rm wave}=5\times10^9~ {\rm \fluxunit}$, we get:

\begin{equation}
    F_{\rm mean} = f_{\rm wave} \times 1.5\times10^6~{\rm \fluxunit}
\end{equation}
which is, for values of $f_{\rm wave}$ around unity, in the range of values given by \citet{Molnar2023wave}. This amount of acoustic flux could, at most, be a contributor to chromospheric heating, which requires, for the low chromosphere alone, about $2\times10^6{\rm \fluxunit}$. Also, the acoustic flux that we have estimated here refers to the upper photosphere and is likely further attenuated up to the temperature minimum. A more accurate estimate of the mean flux requires an analysis of longer observed time series and larger fields of view, and/or the analysis of larger simulated areas with varying initial and boundary conditions. 

From our analysis, we see that this acoustic wave event is strongly correlated to the magnetic field structure, as Fig.\,\ref{fig:wave_detection} shows how the wave perturbation travels alongside the magnetic structures. The magnetic structures decrease the overall pressure in the region of wave propagation (see also Fig.\,\ref{fig:wave_detection}) and hence create chromospheric conditions for 'stronger' wave propagation. 
A preliminary analysis of a similar, purely hydrodynamic, simulation showed no such strong wave events. Purely hydrodynamic, photospheric conditions could possibly weaken such events and make them harder to detect. An in-depth analysis will require more detailed investigation and will be done in a follow-up to this work.

Another complementary aspect would be an analysis of simulations with a realistic chromosphere \citep[e.g.,][]{Przybylski2022muramc, 2026A&A...705A..86N}, to better model the interaction between the waves and the chromospheric layers and identify their role in the heating of the solar chromosphere \citep[see the recent results of ][]{Udnaes_2025_waves}. Ultimately, this study showcases the advantages of high-resolution, high-fidelity IFU spectropolarimetry, combined with realistic rMHD simulations, and their capability to capture the minute details of wave excitation and propagation in the lower solar atmosphere.

\begin{acknowledgement}
We gratefully acknowledge stimulating discussions with J.\,M. Borrero and P. K\"{a}pyl\"{a}. IM acknowledges the financial support from the Serbian Ministry of Science and Technology through the grants 451-03-136/2025-03/200104 and 451-03-136/2025-03/200002.
This research was supported by the Research Council of Norway through its Centres of Excellence scheme, project number 262622.
The Swedish 1-m Solar Telescope is operated on the island of La Palma by the Institute for Solar Physics of Stockholm University in the Spanish Observatorio del Roque de los Muchachos of the Instituto de Astrofísica de Canarias. The Swedish 1-m Solar Telescope, SST, is co-funded by the Swedish Research Council as a national research infrastructure (registration number 4.3-2021-00169).
The simulations were run on Piz Daint at the Swiss National Supercomputing Centre, Switzerland, financed through the ACCESS programme of the SOLARNET project, which has received funding from the European Union Horizon 2020 research and innovation programme under grant agreement no 824135.
This research has made use of NASA's Astrophysics Data System Bibliographic Services. Furthermore we would like to thank the anonymous referee for a constructive review of this work.
\end{acknowledgement}

\bibliography{bibliography}
\end{document}